\documentclass[a4paper, 11pt, leqno, twoside]{amsart}

\usepackage{amssymb}
\usepackage{amsthm}
\usepackage{graphics}
\usepackage{graphicx}

\usepackage[T1]{fontenc}
\usepackage{latexsym}
\usepackage{inputenc}
\usepackage{amsfonts}
\usepackage{amsmath}
\usepackage{amssymb}
\usepackage{pxfonts}
\usepackage{epsfig}
\usepackage{color}
\usepackage{pifont,marvosym}

\newcommand{\pa}{\partial}
\newcommand{\be}{\begin{equation}}
\newcommand{\ee}{\end{equation}}
\newcommand{\bdis}{\begin{displaymath}}
\newcommand{\edis}{\end{displaymath}}
\newcommand{\bit}{\begin{itemize}}
\newcommand{\eit}{\end{itemize}}

\newcommand{\baa}{\begin{array}}
\newcommand{\eaa}{\end{array}}

\begin{document}



\title{A Mathematical model of copper corrosion}


\author{F. Clarelli}
\address{Istituto per le Applicazioni del Calcolo "M. Picone",
Consiglio Nazionale delle Ricerche
Polo Scientifico - Edificio F CNR
Via Madonna del Piano 10; I-50019 Sesto Fiorentino (FI), Italy.}

\author{B. De Filippo}
\address{Istituto per le Applicazioni del Calcolo "M. Picone", Consiglio Nazionale delle Ricerche, via dei Taurini 19, I-00185, Roma, Italy.}

\author{R. Natalini}
\address{Istituto per le Applicazioni del Calcolo "M. Picone", Consiglio Nazionale delle Ricerche, via dei Taurini 19, I-00185, Roma, Italy.}

\begin{abstract} 
A new partial differential model for monitoring and detecting copper corrosion products (mainly brochantite and cuprite)
is proposed to provide predictive tools suitable for describing the evolution of damage induced on bronze specimens by sulfur dioxide ($SO_2$)
pollution. 
This model is characterized by the movement of a double free boundary. Numerical simulations show a nice agreement with experimental result. 
\end{abstract}

\medskip

\subparagraph*{\textup{2000} Mathematics Subject Classification:} 76V05 (35R35, 65M06)

\medskip


\maketitle

\section{Introduction}
\label{par1}

Deterioration of copper and bronze artifacts is one of the main concerns for people working in cultural heritage \cite{Sc02}. More specifically, bronze, a copper-tin alloy, has been widely employed for daily-use and artistic purposes from the Bronze Age up to present. Conservation studies are based on a knowledge of the environmental conditions to which copper and copper alloys may be exposed and include all the information on the material technologies and the nature of the corrosion films or patina, which cover their surfaces. In particular a significant effort has been devoted to study the corrosion due to environmental conditions, such as temperature, moisture, concentration of pollutants \cite{cha90, NMG87, FNS06}. Although in recent years air pollution in European urban areas has decreased considerably, there still remain concentrations of pollutants such as sulfur dioxide ($SO_2$) from combustion of fossil fuels, being one of the most important factors in the deterioration of bronze. Indeed $SO_2$, mixed with water vapor, reacts to produce sulfate acid ($H_2SO_4$), which causes corrosion phenomena on copper surfaces and produces several corrosion products as basic copper sulphates, such as antlerite, posnjakite, brochantite \cite{GNF87}. The latter is the final product of several reaction steps, which can be approximated by two main chemical reactions: cuprite formation, which occurs after a few weeks of exposure to atmospheric conditions, and brochantite formation, which the final reaction step (for more details on chemical background see Section \ref{s21}) \cite{FNA98}. 

The complexity of corrosion processes creates the necessity for a quantitative model approach to develop predictive tools, which simultaneously provide both quantitative information as well as simulations of the various processes involved. These  methods, similar to those introduced in \cite{cla08}, are useful for the monitoring and detection of surface alterations even before they are visible, making it possible to determine optimal intervention strategies. 
In this paper we introduce a new partial differential model, which is used to describe the evolution of damage induced on a bronze specimen by atmospheric pollution. It is based on fluid dynamical and chemical relations and it is characterized by a double free boundary: one between copper and cuprite, the other between cuprite and brochantite. Its calibration has been elaborated according to the experimental results in \cite{def10}. 
The paper is organized as follows: in the Section  \ref{s21} we analyze the main chemical corrosion phenomena and in Section  \ref{s22} we review the main mathematical models already proposed in literature. ...The section \ref{s3} is entirely focused on the description of the model's equations and the numerical schemes used, meanwhile in section \ref{s4} we describe the experimental setting and the calibration of the model. Finally, in section \ref{s5}, we present the main results related to the simulations produced by our model and in section \ref{s6} the related conclusions.

\section{Modeling backgrounds}

\subsection{Chemical backgrounds}\label{s21}
Most of the oxidation processes occurring on bronze metal artifacts, when exposed to environmental conditions, are electrochemical and involve interactions between the metal surface, the adsorbed moisture and various atmospheric gases ($SO_2$, $CO_2$, $NO_x$, hydrocarbons) \cite{pay90}. Electrochemical corrosion processes in electrolyte and condensed moisture layers have been the subject of extensive studies, based on numerous and different approaches \cite{WHI03, SOL06,CBM10}. 
When exposed to the atmosphere, copper and its alloys form a thin layer of corrosion, from brownish-green up to greenish-blue colors, which is designated as patina. In the case of copper in a low pollutant levels atmosphere, a native cuprite (copper(I) oxide or $Cu_2O$) film of approximately a few nanometers thick, protects the metal surface from further oxidation. The general reaction is well described in literature, where, in aerated solution, copper can dissolve electrochemically forming copper(I) oxide formation, due to the reaction of copper with oxygen. It is represented by the following schematic reaction \cite{mac81}: 
$$2Cu + \frac{1}{2}O_2 \rightarrow Cu_2O.$$

However, in an aggressive environment, like urban atmosphere, the protective nature of this oxide layer is altered and there is the formation of a non-protective, multi-component, tarnish layer.
When the copper is exposed to humidity and sulfur dioxide, three kinds of basic copper sulfate hydroxide are mainly produced: Brochantite $Cu_4SO_4(OH)_6$, which is a well-known patina constituent, or other similar products like Antlerite or Posnjakite.
In the following we focus our attention to the formation of Brochantite, which is the main observed product. Gaseous sulfur dioxide and sulfate particles are deposited on the electrolyte on the cuprite. Their deposition reduces the $pH$ of the adsorbed water, and this promotes the dissolution of cuprous ion ($Cu^+$) and its oxidation, thus forming cupric ions ($Cu^{2+}$). In detail, copper(I) ions in solution disproportionate to give copper(II) ions and a precipitate of copper \eqref{eq1}:
\be
\label{eq1}
2Cu^{+}_{(aq)} \to Cu^{2+}_{(aq)} + Cu_{(s)}.
\ee
When the cupric and sulfate ion concentrations in the electrolyte are high enough to form brochantite, this phase starts to precipitate on the cuprite \cite{AWLS00, KOL02}. In \cite{OL95} it is indicated that, in the initial oxidation process, cuprite formation is followed by posnjakite, as a precursor phase to brochantite, see also \cite{WTI01}, but we are going to neglect this intermediate transformations, due to the elevated speed of the reaction with respect to the time scale of the mathematical model which we are going to present.

\subsection{Existing mathematical models}\label{s22}

Graedel and his collaborators \cite{gra96a, tid96} have studied atmospheric copper sulfidation at $AT\&T$ Bell Laboratories in both experimental investigation and physical or mechanistic model development; furthermore a systematic investigation of copper sulfidation kinetics has been performed. In the paper \cite{tid96} Tidblad and Graedel developed a model able to describe the $SO_2$ copper corrosion. Their work is based on aqueous chemistry and without considering spatial dimensions. In the paper \cite{pay95}, they introduce a model which describes the corrosion of copper exposed to moist air with a low $SO_2$ concentration. Here, they consider four stages in the development of a corrosion patina: the metal, a non-protective oxide film which has a high ionic transport property, an outer layer of corrosion products which can permit the penetration of water and gases and an external absorbed water layer. More recently, Larson proposed in \cite{lar02} a different model that describes the atmospheric sulfidation of copper by proposing a physical copper-sulfidation model that includes four distinct phases: the substrate metal, a cohesive cuprous sulfide ($Cu_2S$) product layer, a thin aqueous film adsorbed on the sulphide and the ambient gas. Larson postulated that transport through the sulfide layer occurs via diffusion and electromigration of copper vacancies and electron holes. 
Later, in the $1990$s, copper sulfation by $SO_2$ was investigated by Payer et al., who focused their attention on the early stage of corrosion in moist air ($75\%$ RH at $25^oC$) with a sulfur dioxide concentration of $0.5\%$. The techniques employed (SEM, AES and TEM) allowed for an analysis and characterization of the oxide film (composed of cuprous oxide, copper sulfate and sulfide) on copper surfaces and for the mechanism of evolution of the corrosion chemistry to be described \cite{cha90}. In the last decade the effect of sulfur dioxide in acid rain on copper and bronze has been investigated. Robbiola and his co-workers studied the effect of its cyclic action on bronze alloys, underlining the different types of patina formed in "sheltered" and "unsheltered" areas of bronze monuments \cite{chi06,CBM10}.

Subsequently, Larson refined the copper-sulfidation model by focusing on the transport of charged lattice defects in a growing $Cu_2S$ product layer between the ambient gas and the substrate metal. As previously, this transport is postulated to occur via both diffusion and electromigration.

\section{The model}
\label{s3}

\begin{figure}[!ht]
    \begin{center}
        \includegraphics[width=10cm, height=7cm]{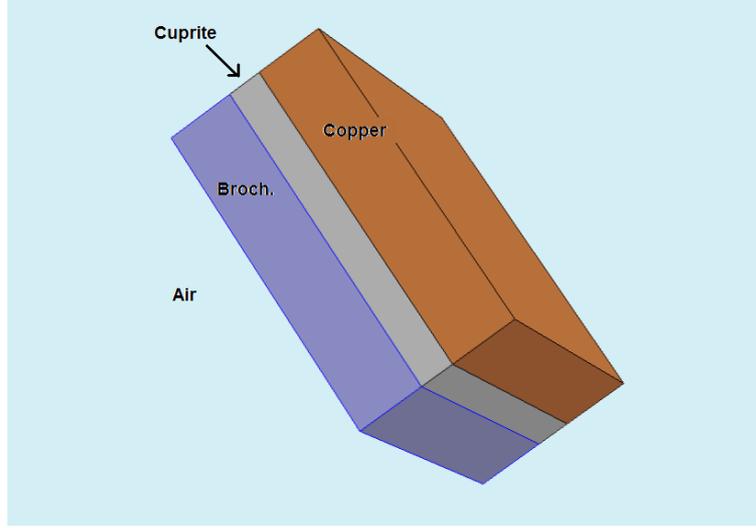}
    \end{center}
    \caption{\footnotesize{Example of cuprite and brochantite deposition on a copper sample.}}
    \label{fig1}
\end{figure}

In this section, we aim to introduce a mathematical model able to describe the corrosion effects on a copper layer, which is subject to deposition of $SO_2$. The present model is based on the mathematical approach used in \cite{cla08}. We assume to have a copper sample on which is formed a non protective oxide layer ($Cu_2O$), and, over this layer, a corrosion product (brochantite) grows. Over the brochantite layer is assumed to be the atmospheric air with $SO_2$. An example of these three layers can be seen in Figure \ref{fig1}.

  The reaction producing cuprite $Cu_2O$, can be approximated by 
\begin{equation}
\label{eq1bis}
2Cu + \frac{1}{2}O_2 \rightarrow Cu_2O.
\end{equation}

Namely, two moles of copper combined with one-half mole of oxygen produce cuprite.

Laboratory tests, with high concentration of $SO_2$ and high relative humidity, near to $100 \%$, show that brochantite is the primary product of the reaction, thus it can be considered as the final state reached by the whole reaction.

The overall simplified reaction of brochantite formation can be approximate by the following reaction (\ref{eq2}) 
\begin{equation}
\label{eq2} 
2 Cu_2O + SO_2 + 3H_2O + \frac{3}{2}O_2 \rightarrow Cu_4 (OH)_6 SO_4,
\end{equation}
where two moles of cuprite combined with three moles of water and three-halves of oxygen produce one mole of brochantite.

In the following, we assume that these two reactions are instantaneous; thus, we obtain a sharp free boundary between cuprite and the unreacted copper, due to the reaction (\ref{eq1bis}), and a second free boundary between brochantite and cuprite, due to the reaction (\ref{eq2}). 

Assuming these two reactions as instantaneous, the effective time of reaction is implicitly included in the diffusivity coefficients.

\subsection{Swelling}

We indicate the copper consumption by $a(t)$. The production of $Cu_2O$ on the boundary between copper and the oxide layer proceeds since the water and the oxygen diffuse through the oxide layer. On the upper boundary of cuprous oxide, a brochantite layer begins to form. It is assumed that, in the climatic chamber at a $100\%$ of RH, a thin film of water (with $SO_2$ dissolved) is formed over the cuprous oxide, and it plays an important role in both reactions. 
By these assumptions, the consumption of $Cu_2O$ and the production of brochantite on the upper boundary between oxide and water film occurs. The volume of cuprite consumption is $b(t)$.

The transformation of copper into $Cu_2O$, such as the transformation of cuprite in brochantite, are accompanied by a volume change 
(swelling rate). The swelling rate can be calculated easily, because the molar ratio in reaction (\ref{eq1}) between $Cu$ and $Cu_2O$ is $2 : 1$. Thus two moles of copper change into one mole of $Cu_2O$, and a different volume of the new matter formed is obtained. 
The swelling of reaction (\ref{eq1}) is
\be
\label{eq3}
\dot{a_e} = - \omega_p \dot{a};
\ee
where $a_e$ is the swelling of reaction (\ref{eq1}), and 
\be
\label{eq3a}
\omega_p = \frac{\mu_c}{2 \mu_p} - 1
\ee
represents the expansion volume ratio; $\mu_c$ and $\mu_p$ being the molar density $(moles/cm^3)$ of $Cu$ and $Cu_2O$ respectively. It is assumed that $\mu_c$ and that $\mu_p$ are constant (i.e. they are homogeneous materials). Under these assumptions, if $a_e (0) = a(0) = 0$, then it is possible to conclude that 
\be
\label{eq4}
a_e(t) = - \omega_p a(t),
\ee
and the thickness of cuprite layer $h_p(t)$, is proportional to the copper consumption $a(t)$
\be
\label{eq5}
h_p(t) = \left( 1 + \omega_p \right) a(t).
\ee

On the external boundary of cuprite, we assume that $SO_2$ reacts with $Cu_2O$ in presence of water, see reaction (\ref{eq2}). Here, we have that 2 moles of cuprite change in 1 mole of brochantite, wasting 1 mole of $SO_2$, 3 moles of $H_2O$ and $3/2$ moles of oxygen. Thus, the brochantite layer grows on the external boundary of cuprite, and we indicate the external boundary of brochantite with $\gamma(t)$.

As in equation (\ref{eq4}), the swelling rate of brochantite is given by
\be
\label{eq4a}
\dot{b}_e(t) = - \omega_b \dot{b}(t);
\ee
where $b_e(t)$ is the swelling of reaction (\ref{eq2}),
\be
\label{eq5a}
\omega_b = \frac{\mu_p}{2 \mu_b} - 1,
\ee
and $\mu_b $ is the constant molar density of brochantite. 

The cuprite consumption $b(t)$ is referred to a system of reference at rest, but we have to take into account the moving cuprite boundary. The physical boundary between cuprite and brochantite is given by 
\be
\label{eq5b}
\beta(t) = b(t) + a_e(t), 
\ee
where $b(t)$ is the consumption of cuprite (assumed to be positive such as $a(t)$), then $a_e<0$, see eq. (\ref{eq4}). Thus, the variation of the overall external boundary $\dot{\gamma}$ is given by
\be
\label{eq4b}
\dot{\gamma} = - \omega_p \dot{a}(t) - \omega_b \dot{b}(t).
\ee

\begin{figure}[!ht]
    \begin{center}
        \includegraphics[width=10cm, height=5cm]{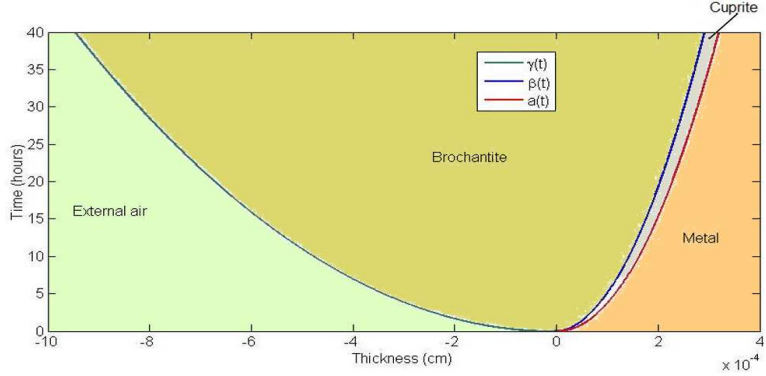}
    \end{center}
    \caption{\footnotesize{Example of  cuprite and brochantite theoretical growth (in time) on a copper sample.}}
    \label{fig2}
\end{figure}

Summarizing, the geometry of our problem is one-dimensional, and we have $4$ regions (e.g. see figure \ref{fig2}):
\begin{enumerate}
	\item Copper (inner region).
	\item Cuprite $Cu_2O$, between $a(t)$ and $\beta(t)$.
	\item Brochantite $Cu_4 (OH)_6 SO_4$, between $\beta(t)$ and $\gamma(t)$.
	\item Water film and air on the external side of $\gamma(t)$.
\end{enumerate}

\subsection{Equation of the model}

It is well known that relative humidity plays a key role in regulating the speed of oxydation and sulfation. It has been observed that when relative humidity exceeds some threshold, then $SO_2$ reacts completely such as the oxidation of copper happens with full speed. We can interpret this phenomenon as follows.
According to eq. (\ref{eq2}), when a molecule of $SO_2$ comes in contact with cuprite, it reacts if three molecules of $H_2O$ are available at the same point (we suppose that there is always enough $O_2$). To be more precise, this is true only if vapor condenses on the unreacted specimen surface forming a liquid film, and this happens with high relative humidity values.

The brochantite formation (eq. (\ref{eq2})) has been assumed to develop on the cuprite layer, due to the $SO_2$, $H_2O$ and $O_2$, which move through the brochantite layer and react with $Cu_2O$. This reaction implies a wasting of $Cu_2O$, with the formation of a new layer of brochantite. Summarizing, the region of brochantite is given by $\gamma(t) \leq x \leq \beta(t)$ and the region of $Cu_2O$ is $\beta(t) \leq x \leq a$.

Let's denote the concentration of $SO_2$ in the pores of brochantite by $S$, coming from the external air, such as the water concentration indicated by $W$ and the oxygen concentration by $O$ throughout brochantite and by $G$ throughout cuprite.

The flow of $SO_2$ relative to air is governed by Fick's law. Thus, in the frame of reference where copper is at rest, the $SO_2$ flux has the following expression 
\be
\label{eq6}
J_s = n_b \left( - D_s \frac{\partial S}{\partial x} - S \omega_p \dot{a} - S \omega_b \dot{b} \right) = n_b \left( - D_s \frac{\partial S}{\partial x} + S \dot{\gamma} \right),
\ee
where the first term refers to $SO_2$ diffusion in the brochantite layer, $D_s$ is the diffusivity, the second and the third terms on the left side refer to the swelling caused by cuprite and brochantite formation respectively.

Hence the mass balance of $SO_2$ in the brochantite layer $\gamma(t) \leq x \leq \beta(t)$ is 
\be
\label{eq6a}
\frac{\partial S}{\partial t} - D_s \frac{\partial^2 S}{\partial x^2} +  \dot{\gamma} \frac{\partial S}{\partial x}  = 0;
\ee
The value of $S$ at the external boundary $\gamma(t)$ is the environment $SO_2$ concentration $S_a(t)$, which is a known function of time
\be
\label{eq6b}
S(\gamma(t)) = S_a(t);
\ee
We assumed that $SO_2$ reacts totally with $Cu_2O$ at the front $\beta(t)$, thus we have
\be
\label{eq6c}
S(\beta(t)) = 0.
\ee
Now, we need of further condition (first free boundary). Since the $SO_2$ flux of moles at the boundary $\beta(t)$ is proportional to $Cu_2O$ moles consumption, we have
\be
\label{eq6d}
- n_b \frac{D_s}{M_s} \frac{\partial S}{\partial x} = \frac{1}{2} \frac{\rho_p}{M_p} \dot{b}; 
\ee
where $M_s$, $M_p$ are the molar weight of $SO_2$ and $Cu_2O$ respectively, $\rho_p$ is the mass density of cuprite.

Similarly, the water flux is
\be
\label{eq7}
J_w = n_b \left( - D_w \frac{\partial W}{\partial x} - W \omega_p \dot{a} - W \omega_b \dot{b} \right) = n_b \left( - D_w \frac{\partial W}{\partial x} + W \dot{\gamma} \right),
\ee
where $D_w$ is the water diffusivity. Thus, the mass balance in the brochantite layer is 
\be
\label{eq7a}
\frac{\partial W}{\partial t} - D_w \frac{\partial^2 W}{\partial x^2} + \dot{\gamma}  \frac{\partial W}{\partial x}  = 0.
\ee
The value of $W$ at the external boundary $\gamma(t)$ is the environment water concentration $W_a(t)$
\be
\label{eq7b}
W(\gamma(t)) = W_a(t);
\ee
Since some moles of water are wasted by the reaction (\ref{eq2}), at the front $\beta(t)$ we have
\be
\label{eq7c}
\frac{J_w}{M_w} = \frac{3}{2} \frac{\rho_p}{M_p} \dot{b} + n_b \frac{W}{M_w} \dot{b}.
\ee

Finally, the oxygen flux is
\be
\label{e1}
J_o = n_b \left( - D_o \frac{\partial O}{\partial x} + O \dot{\gamma} \right),
\ee
where $D_o$ is the oxygen diffusivity. Thus, the mass balance in the brochantite layer is 
\be
\label{e2}
\frac{\partial O}{\partial t} - D_o \frac{\partial^2 O}{\partial x^2} + \dot{\gamma}  \frac{\partial O}{\partial x}  = 0.
\ee
The value of $O$ at the external boundary $\gamma(t)$ is the environment oxygen concentration $O_a(t)$
\be
\label{e3}
O(\gamma(t)) = O_a(t).
\ee
Since the oxygen is also wasted by the reaction (\ref{eq2}), at the front $\beta(t)$ we have
\be
\label{e4}
\frac{J_o}{M_o} = \frac{3}{4} \frac{\rho_p}{M_p} \dot{b} + n_b \frac{O}{M_o} \dot{b}.
\ee

\subsection{Cuprite layer equations}

We assumed that the reaction between oxygen and copper occurs on the
copper-cuprous oxide boundary $\alpha(t)$. Here, we indicate oxygen by $G$, just to avoid confusion with the previous region. We make a mass balance of oxygen concentration $G$, and we assume that all oxygen moles arriving on the inner boundary $a(t)$ react. This assumption is not really true, in fact the water plays a key role in the speed of reaction, but in our experiments we have a relative humidity near to $100 \%$, and this fact justify our assumption; also, the diffusivity $D_g$ includes implicitly the finite time of reaction.  

The flux of oxygen is
\be
\label{eq8}
J_g = n_p \left( - D_g \frac{\partial G}{\partial x} - G \omega_p \dot{a} \right),
\ee
and the mass balance equation is
\be
\label{eq9}
\frac{\partial G}{\partial t} - D_g \frac{\partial^2 G}{\partial x^2} - \omega_p \dot{a} \frac{\partial G}{\partial x}  = 0.
\ee

The value of $G$ at the boundary $\beta(t)$ is given by the value of oxygen on the boundary $\beta$, given by eq. (\ref{e4}).
\be
\label{eq9a}
G(\beta(t)) = O(\beta(t));
\ee
Since oxygen reacts totally with copper at the boundary $a(t)$, we have
\be
\label{eq9b}
G(a(t)) = 0;
\ee
Now, we need further condition to close the system (second free boundary). Since the oxygen react totally at the boundary $\alpha(t)$, the copper moles wasted are given by 
\be
\label{eq9c}
- n_p \frac{D_g}{M_g} \frac{\partial G}{\partial x} = \frac{1}{4} \frac{\rho_c}{M_c} \dot{a}. 
\ee

\subsection{Rescaling}

The geometry of the problem is given by two regions, one is given by $x \in [\beta(t),a(t)]$ which describes the layer of $Cu_2O$, the other is given by $x \in [\gamma(t),\beta(t)]$ i.e. the brochantite layer. Also we have $ \dot{\gamma} = - \left( \omega_p \dot{a} + \omega_b \dot{b} \right)$.

\subsubsection{New variables}

In the first region we adopt $(x,t) \rightarrow (y,\tau)$, in the second one $(x,t) \rightarrow (z,\tau)$. We find out

\begin{subequations}
\begin{align}
    y =& \frac{x - \beta}{a - \beta};\; \partial_x = 1/(a - \beta)\partial_y, \; \partial_{xx} = 1/(a - \beta)^2\partial_{yy},\; 
    \pa_t = f(t)\pa_y +\frac{1}{t_r}\pa_{\tau};
   \label{res1} \\
    z =& \frac{x - \gamma}{\beta - \gamma};\; \partial_x = 1/(\beta - \gamma)\partial_z, \; \partial_{xx} = 1/(\beta - \gamma)^2 \partial_{zz},\;
    \pa_t = q(t)\pa_z + \frac{1}{t_r}\pa_{\tau}; 
  \label{res2} 
\end{align}
\label{res}
\end{subequations}
where
\begin{subequations}
\begin{align}
    f(y,\tau) =& \frac{1}{t_r} \frac{y ( \pa_{\tau} \beta - \pa_{\tau}a ) - \pa_{\tau}\beta }{a - \beta},
   \label{resa1} \\
    q(z,\tau) =& \frac{1}{t_r} \frac{z (\pa_{\tau}\gamma - \pa_{\tau}\beta) - \pa_{\tau}\gamma}{ \beta - \gamma}. 
  \label{resa2} 
\end{align}
\label{resa}
\end{subequations}

Now, we rescale the following parameters and variables in non-dimensional form:
\begin{subequations}
\begin{align}
		\hat{W} =& \frac{W}{W_r}, \; \hat{G} = \frac{G}{G_r}, \; \hat{S} = \frac{S}{S_r}, \; \hat{a} = \frac{a}{\lambda}, \; \hat{b} = \frac{b}{\lambda}, \; \hat{\beta} = \frac{\beta}{\lambda}, \; \hat{\gamma} = \frac{\gamma}{\lambda},
		\label{ndim1}\\
    \hat{D}_{g} =& \frac{t_r}{\lambda^2} D_{g},\;\hat{D}_{w} = \frac{t_r}{\lambda^2} D_w,\;\hat{D}_s = \frac{t_r}{\lambda^2} D_s,\; \hat{q}(z,\tau) = t_r q, \; \hat{f}(y,\tau) = t_r f 
    \label{ndim2}
\end{align}
\label{ndim}
\end{subequations}

From now on, we assume that dotted variables are derivatives with respect to the dimensionless time $\tau$ $(\dot{V} = dV/d\tau )$.

Assuming $\Omega_s = 2 n_b \hat{D}_s \frac{M_p}{M_s} \frac{S_r}{\rho_p}$, $\Gamma_w = \frac{3}{2} \frac{1}{n_b} \frac{M_w}{M_p} \frac{\rho_p}{W_r}$, $\Omega_g = 4 n_p \hat{D}_{g} \frac{M_c}{M_g} \frac{G_r}{\rho_c}$. Also, $\dot{\hat{\gamma}} = - ( \omega_p \dot{\hat{a}} + \omega_b \dot{\hat{b}} )$, the non-dimensional system is:

\subsubsection{Outer region}

For $\gamma \leq x \leq \beta \rightarrow 0 \leq z \leq 1$

\be 
\label{nd5}
\frac{\pa \hat{S}}{\pa \tau} = \frac{\hat{D}_s}{\left(\hat{\beta} - \hat{\gamma} \right)^2} \frac{\pa^{2} \hat{S}}{\pa z^{2}} - \left( \frac{  \dot{\hat{\gamma}} }{\left(\hat{\beta} - \hat{\gamma} \right)} + \hat{q} \right) \frac{\pa \hat{S}}{\pa z},
\ee

\be
\label{nd6}
\hat{S}(0,\tau) = \hat{S}_a.
\ee

\be
\label{nd7}
\hat{S}(1,\tau)=0,
\ee

\be 
\label{nd8}
- \frac{\Omega_s}{\left( \hat{\beta} - \hat{\gamma} \right)}\frac{\pa \hat{S}}{\pa z}(1,\tau) = \dot{\hat{b}},
\ee

\be 
\label{nd9}
\frac{\pa \hat{W}}{\pa \tau} = \frac{\hat{D}_w}{\left(\hat{\beta} - \hat{\gamma} \right)^2} \frac{\pa^{2} \hat{W}}{\pa z^{2}} - \left( \frac{  \dot{\hat{\gamma}} }{\left(\hat{\beta} - \hat{\gamma} \right)} + \hat{q} \right) \frac{\pa \hat{W}}{\pa z},
\ee

\be
\label{nd10}
\hat{W}(0,\tau)=\hat{W}_a(\tau).
\ee

\be
\label{nd11}
\frac{\hat{D}_w}{\left(\hat{\beta} - \hat{\gamma} \right)} \frac{\pa \hat{W}}{\pa z}(1,\tau) = \left( \dot{\gamma} - \dot{b} \right) \hat{W} - \frac{3}{2 n_b}\frac{\rho_p}{W_r}\frac{M_w}{M_p}\dot{b};
\ee

\be 
\label{nd9a}
\frac{\pa \hat{O}}{\pa \tau} = \frac{\hat{D}_o}{\left(\hat{\beta} - \hat{\gamma} \right)^2} \frac{\pa^{2} \hat{O}}{\pa z^{2}} - \left( \frac{  \dot{\hat{\gamma}} }{\left(\hat{\beta} - \hat{\gamma} \right)} + \hat{q} \right) \frac{\pa \hat{O}}{\pa z},
\ee

\be
\label{nd10a}
\hat{O}(0,\tau)=\hat{O}_a(\tau).
\ee

\be
\label{nd11a}
\frac{\hat{D}_o}{\left(\hat{\beta} - \hat{\gamma} \right)} \frac{\pa \hat{O}}{\pa z}(1,\tau) = \left( \dot{\gamma} - \dot{b} \right) \hat{O} - \frac{3}{4 n_b}\frac{\rho_p}{O_r}\frac{M_o}{M_p}\dot{b},
\ee

\subsubsection{Inner region}

For $\beta \leq x \leq a \rightarrow 0 \leq y \leq 1$,
\be
\label{nd1}  
\frac{\pa \hat{G}}{\pa \tau} = \frac{\hat{D}_{g}}{ (\hat{a} - \hat{\beta} )^2} \frac{\pa^2 \hat{G}}{\pa y^2} + \left( \omega_p \frac{\dot{\hat{a}}}{\hat{a} - \hat{\beta}} - \hat{f}(y,\tau) \right) \frac{\pa \hat{G}}{\pa y},
\ee
(we have to highlight that $\dot{a}$ is a derivative with respect to the $\tau$)
\be
\label{nd2}  
\hat{G}(1,\tau) = 0,
\ee
\be
\label{nd3}  
\hat{G}(0,\tau) = \left. \hat{O}\right|_{\beta},
\ee
\be
\label{nd4}  
- \frac{\Omega_w}{\hat{a} - \hat{\beta}} \frac{\pa \hat{G}}{\pa y} = \dot{\hat{a}}.
\ee
From now on we use the non-dimensional system, so we indicate all non-dimensional terms without hat.

\subsection{Numerical solutions}

To solve our model, we have to set up an appropriate numerical scheme which is able to describe the process in a short range of time (about $40$ hours), but also simulations of $1$ year taking under control the numerical stability. For these reasons we use finite differences schemes with implicit-explicit terms. It can be found in \cite{bri07} and \cite{ruu95}. 

\subsubsection{Initial conditions} 
Our numerical procedure requires $a(0) \neq 0$ and $\beta(0) \neq a(0)$ to avoid singularities in the internal region confining with copper. By physical assumption, we know that $a(0) > 0$ because it represents the first copper consumption. In order that the external equations work ($SO_2$ is present), we need to set up $\beta(0) > 0$ and $\gamma(0) \neq \beta(0)$.

\subsubsection{Numerical scheme}

Indicating explicit term as $H(U)$ and implicit term by $G(U)$, we present our system in the external region (eqs. \ref{nd5}-\ref{nd11}) in the following form 
\be
\label{ns1}  
U_t = H(U) + G(U);
\ee
where 
\begin{center}
\begin{math}
U = \left( 
\begin{array}{c}
S \\       
 W \\
  O 
\end{array}
\right),
\end{math}
\end{center}
the explicit term $H(U)$ is
\begin{center}
\begin{math}
H(U) = \left( 
\begin{array}{cc}
  \displaystyle - \left( \frac{  \dot{\gamma} }{\left(\beta - \gamma \right)} + q \right) \frac{\pa S}{\pa z} \\  
  \displaystyle - \left( \frac{  \dot{\gamma} }{\left(\beta - \gamma \right)} + q \right) \frac{\pa W}{\pa z} \\
  \displaystyle - \left( \frac{  \dot{\gamma} }{\left(\beta - \gamma \right)} + q \right) \frac{\pa O}{\pa z}
\end{array}
\right),
\end{math}
\end{center}
and the implicit term $G(U)$ is
\begin{center}
\begin{math}
G(U) = \left( 
\begin{array}{cc}
\displaystyle \frac{D_s}{\left(\beta - \gamma \right)^2} \frac{\pa^{2} S}{\pa z^{2}} \\  
 \displaystyle \frac{D_w}{\left(\beta - \gamma \right)^2} \frac{\pa^{2} W}{\pa z^{2}} \\
  \displaystyle \frac{D_o}{\left(\beta - \gamma \right)^2} \frac{\pa^{2} O}{\pa z^{2}}
\end{array}
\right),
\end{math}
\end{center}
since $G(U)$ is a stiff term which will be integrated implicitly to avoid excessively small time steps.

A more general scheme is IMEX-DIRK Runge-Kutta, which is given by, for $t = n \Delta t$,
\be
\label{3im7}
u^{(i)} = u^n + \Delta t \sum_{k=1}^{i-1} \tilde{a}_{ik}H(u^{(k)}) + \Delta t \sum_{k=1}^{i} a_{ik} G(u^{(k)}) ,~~i =1,... \nu 
\ee
\be
\label{3im8}
u^{n+1} = u^n + \Delta t \sum_{i = 1}^{\nu} \tilde{\omega}_{i} H(u^{(i)}) + \Delta t \sum_{i = 1}^{\nu} \omega_{i} G(u^{(i)}).
\ee

Here, the matrices $\tilde{A} = (\tilde{a}_{ik})$, where $\tilde{a}_{ik} = 0$ for $j \geq i$ and $A = (a_{ik})$ are $\nu \times \nu $ matrices such that the resulting scheme is explicit in H and implicit in G. The DIRK formulation requires $a_{ik} = 0$ for $j > i$ \cite{bri07}. 

Following the IMEX formalism we shall use the notation Name($s$; $\nu$; $p$) to identify a scheme, where $s$ is the number of stages of the implicit scheme, $\nu$ is the number of explicit stages and $p$ is the combined order of the scheme.
In our case we choose to use an Implicit-Explicit Midpoint($1$,$2$,$2$): $s = 1$, $\nu = 2$ and $p=2$.

By the same method, and at the same time-step of previous integration, we integrated also equations \eqref{nd1}-\eqref{nd4}.

\section{Assessment of the model}
\label{s4}

\begin{table}[ht!]
\caption{Parameters table}
\centering
\label{param1}       
\begin{tabular}{|llll|}
\hline\noalign{\smallskip}
Param. & Value & Dimension & Indications  \\[3pt]
\hline
\noalign{\smallskip}
$\rho_p$ & $6.00$ & $g/cm^3$ & $Cu_2O$  mass density \\
$M_p$ & $143.09$ & $ g/mol$ & $Cu_2O$  molar mass \\
$\rho_c$ & $8.94$ & $g/cm^3$ & Copper  mass density \\
$M_c$ & $63.55$ & $g/mol$ & Copper  molar mass \\
$\rho_b$ & $3.97$ &$g/cm^3$ & Brochantite mass density \\
$M_b$ & $452.3$ & $g/mol$ & Brochantite molar mass \\
$\rho_s$ & $1.46$ & $g/cm^3$ & $SO_2$ mass density \\
$M_s$ & $64.07$ & $g/mol$ & $SO_2$ molar mass \\
\noalign{\smallskip}\hline
\end{tabular}
\end{table}

\subsection{Experiments}
          
The calibration of the model has been made with reference to a precise experimental campaign, which is described in \cite{def10}, where the main characteristics of the experimental setting are reported. Experiments were carried out in a cyclic corrosion cabinet (Erchisen Mod. 519/AUTO) and the $Cu-12Sn$ cast bronze specimens were chosen as a representative alloy used in the past (ancient Greek type). For the corrosion test, all bronze specimens were exposed to an atmosphere containing about $200$ $ppm$ of $SO_2$ at $40^o C$ and $100\%$ RH for 8 hours (wet cycle), subsequently they were exposed to room conditions for 16 hours (dry cycle). Each wet and dry cycle was repeated 20 times. 
The measures of patina thickness, useful for the model calibration, were performed using a SEM-EDS instrument equipped with an Image Analyzer (IA). Each measurement (after 8, 24 and 40 hours) has been obtained by an average of 20 measures.

\begin{figure}[!ht]
    \begin{center}
        \includegraphics[width=9cm, height=7cm]{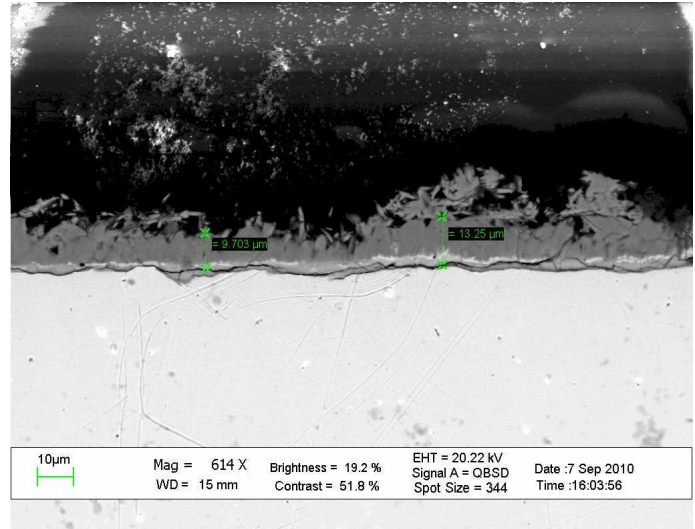}
    \end{center}
    \caption{\footnotesize{Example of patina thickness measurement with the SEM employment. From \cite{def10}}}
    \label{I0}
\end{figure}

The early stage of exposure of bronze samples mainly produced copper hydroxyl-sulfate (brochantite and chalcanthite), cuprous oxide, and tin sulfide (ottemannite). The tarnish product layer also contained a trace of tin oxide. 
As determined by XRD analysis, the brochantite formation was predominant with the increase of the number of corrosion cycles. Otherwise, the chalcanthite formation was detected during the early stage of the corrosion processes and the intensity of its characteristic peaks decreases as a function of time. The same consideration can be made for the ottemannite formation, detected with SEM-EDS in localized micro-crack areas, which are rapidly covered by basic copper sulfates \cite{def10, def11}.

\subsection{Calibration}

The laboratory corrosion tests has been used to calibrate the model.
We suppose that the parameters to calibrate by experimental tests are the diffusivity coefficients. To do that, we used the thickness of corrosion products measured in some time points. Each thickness value has been obtained by an average of $20$ measures, see figure \ref{thick} for two measures of them, and we have obtained the following values:

\begin{table}[htp]
\caption{Patina thickness measures.}
{\begin{tabular}{c|c|c} \hline
Time of measure (hours) & Averaged value $(cm)$ & Standard deviation  \\ \hline
$8$  & $5.4418  \cdot10^{-4} $ & $1.7331 \cdot10^{-4} $  \\
$24$ & $9.2672  \cdot10^{-4} $ & $1.8473 \cdot10^{-4} $  \\
$40$ & $13.2522 \cdot 10^{-4}$ & $2.4102 \cdot10^{-4} $  \\ \hline
\end{tabular}}
\label{thick}
\end{table}

\begin{figure}[!ht]
    \begin{center}
        \includegraphics[width=10cm, height=9cm]{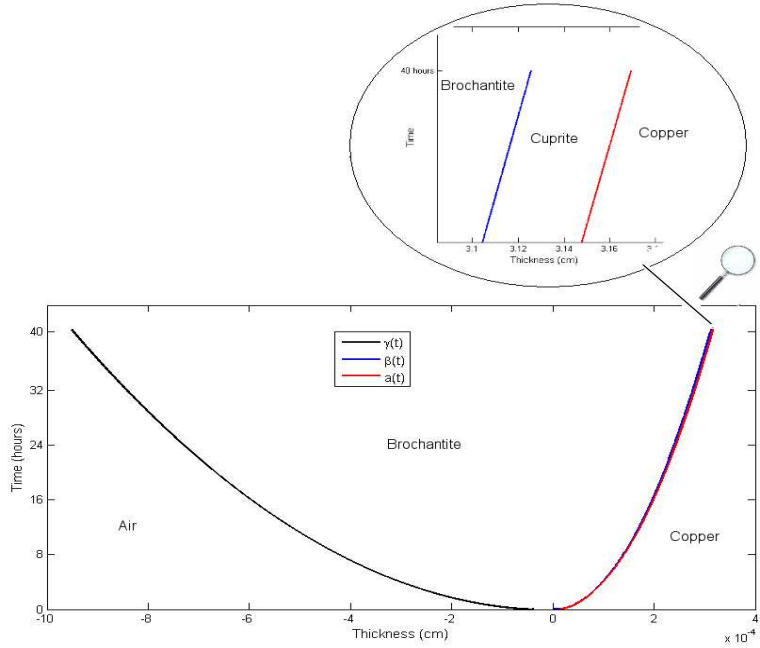}
    \end{center}
    \caption{\footnotesize{Evolution of $\gamma(t)$, $\beta(t)$ and $a(t)$.}}
    \label{curves}
\end{figure}

\begin{figure}[!ht]
    \begin{center}
        \includegraphics[width=12cm, height=6cm]{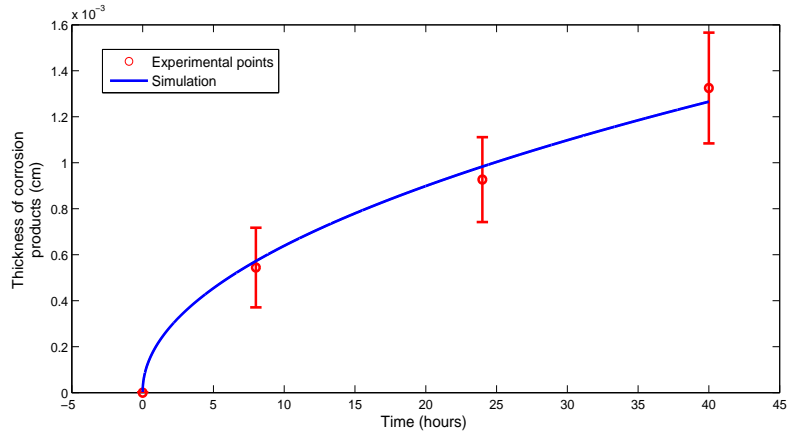}
    \end{center}
    \caption{\footnotesize{Simulation (continuous line) and experimental points (in red).}}
    \label{comp}
\end{figure}

Then, we used the least square method to find the best parameters. We obtained (in $cm^2/sec$): $D_g = 9.9 \cdot 10^{-9}$, $D_s = 3.96 \cdot 10^{-5}$, $D_o = 9.9 \cdot 10^{-6}$ and $D_w= 3.96 \cdot 10^{-5}$. The evolution in time of the corrosion products thickness is in figure \ref{curves}. 
We can see the difference between the experimental points and the best simulation in figure \ref{comp}

{\bf{Remark}}
Here, we test numerical scheme to reproduce the chemical reactions in eqs. (\ref{eq1}) and (\ref{eq2}). Simulations, after 40 hours, give us the following values: $\gamma = -9.505 \cdot 10^{-4}$ $(cm)$, $a = 3.1693 \cdot 10^{-4}$ $(cm)$, $b = 7.9916 \cdot 10^{-4}$ $(cm)$. 
Knowing the molar volume of copper $V_c = \rho_c/M_c$, cuprite $V_p=\rho_p/M_p$, and brochantite $V_b = \rho_b/M_b$, we find that the number of copper moles wasted are $a/V_c = 4.4588 \cdot 10^{-5}$, the number of cuprite moles formed are $h_p/V_p = 2.2294 \cdot 10^{-5}$, i.e. two copper moles are wasted to form one cuprite mole, as we expect from eq. (\ref{eq1}). Similarly, we obtain the number of cuprite moles wasted by reaction (\ref{eq2}) is $b/V_p = 2.2173 \cdot 10^{-5}$, while the number of brochantite moles formed by the same reaction is $h_b/V_b = 1.1086 \cdot 10^{-5}$, i.e. for each brochantite mole formed two cuprite moles are wasted, as expected from (\ref{eq2}).

\section{Application}
\label{s5}

\begin{figure}[!ht]
    \begin{center}
        \includegraphics[width=10cm, height=6cm]{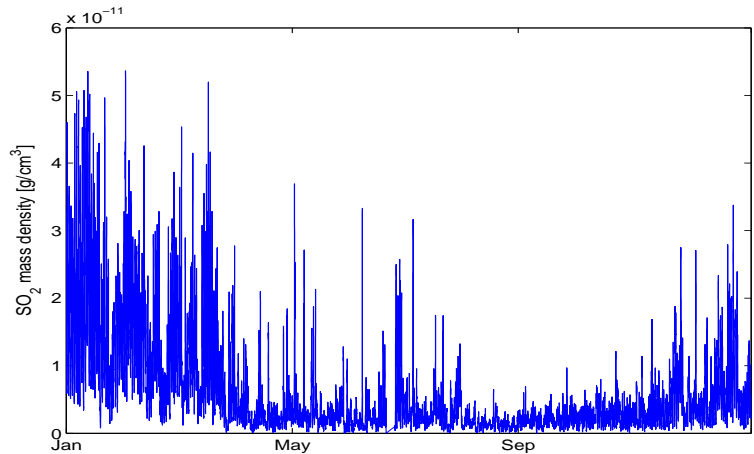}
    \end{center}
    \caption{\footnotesize{$SO_2$ concentration detected at Piazzale Fermi (Rome, Italy) during the 2005.}}
    \label{fso2}
\end{figure}

\begin{figure}[!ht]
    \begin{center}
        \includegraphics[width=10cm, height=6cm]{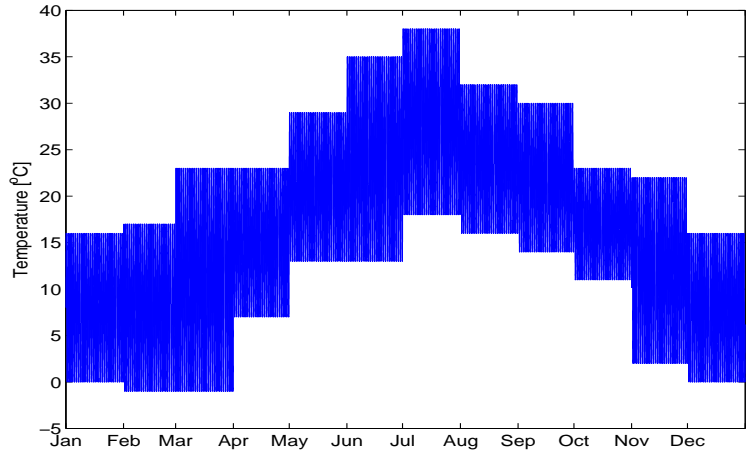}
    \end{center}
    \caption{\footnotesize{Temperature detected at Piazzale Fermi (Rome, Italy) during the 2005.}}
    \label{ft}
\end{figure}

In this section we assume to use the calibrated model with environmental data, detected in Rome at Piazzale Fermi during the year $2005$. This application is just an example on how the model can be used. Since the model has been calibrated for high relative humidity and $SO_2$ concentration, the calibration should be improved by experiments with lower values of relative humidity and $SO_2$ concentration. 
The $SO_2$ concentration detected is in figure \ref{fso2}
and the temperature is in figure \ref{ft}.

\begin{figure}[!ht]
    \begin{center}
        \includegraphics[width=12cm, height=6cm]{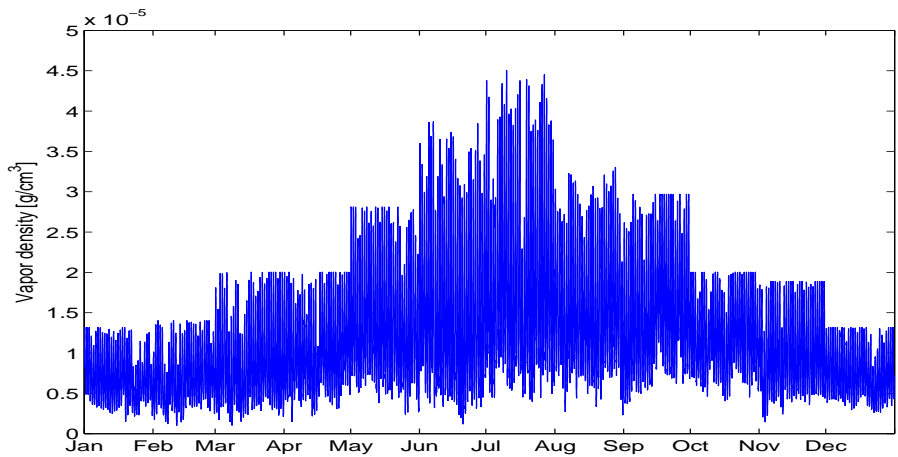}
    \end{center}
    \caption{\footnotesize{Saturated vapor density obtained by temperature and relative humidity detected at Piazzale Fermi (Rome, Italy).}}
    \label{fsvd}
\end{figure}

Assuming to have a good calibration with our environmental data, we need of saturated vapor density ($SVD$ in $[g/cm^3]$) as function of relative humidity and temperature $T$ $[^oC]$ detected at Piazzale Fermi. It is useful for getting an exact quantity of water vapor in the air from a relative humidity (RH), the density of water in the air is given by $RH \cdot SVD = Actual Vapor Density$. The behavior of water vapor density is a non-linear function, but an approximate calculation of saturated vapor density can be made from an empirical fit of the vapor density curve \ref{svd} (between $0-45$ $^oC$ is a good approximation).
\be
\label{svd}  
SVD(T) = 5.018 + 0.32321 T + 8.1847 \cdot 10^{-3} T^2 + 3.1243 \cdot 10^{-4} T^3.
\ee
The figure of saturated vapor density obtained is in figure \ref{fsvd}

\begin{figure}[!ht]
    \begin{center}
        \includegraphics[width=12cm, height=6cm]{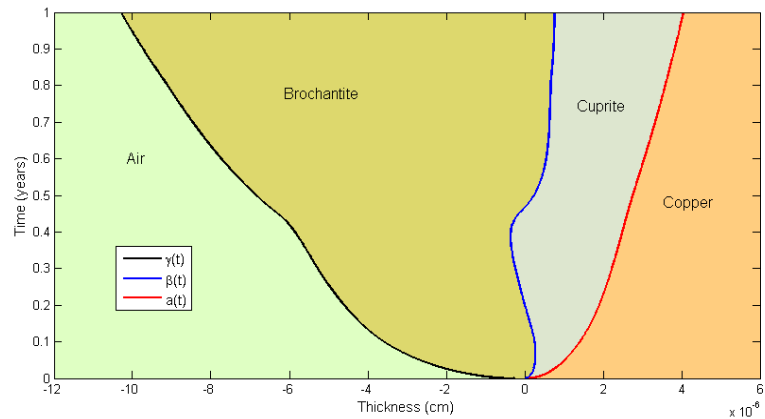}
    \end{center}
    \caption{\footnotesize{Simulation of corrosion with environmental data detected at Piazzale Fermi (Rome, Italy) in 12 months (2005).}}
    \label{rome05}
\end{figure}

Thus, using $SO_2$, temperature and humidity data detected at Piazzale Fermi in Rome during the 2005, we obtained the behavior of corrosion products illustrated in figure \ref{rome05}.

Results of this simulation represent a qualitative example of corrosion formation on copper under $SO_2$ attack. They have to be improved by others laboratory experiments, because our model is calibrated using a very high $SO_2$ concentration (never present in the atmospheric environment in this concentration), and under a very high relative humidity (near to $100 \%$). To have a more accurate calibration, it would be necessary set up mainly experiments with different $SO_2$ concentration and different concentration of humidity in the air, so to calibrate the model under these variations. 

Our conditions produce a very thick cuprite layer in 1-year simulations with respect to the laboratory experiments, because of the air $SO_2$ concentration is very lower than that present in the experimental room, and this caused a slower growth of brochantite layer.  

\newpage
\section{Conclusions}
\label{s6}

In conclusion a new mathematical model was developed to describe and simulate the evolution of brochantite formation on Cultural Heritage artifacts exposed to sulfur dioxide atmosphere. The aim was to create a new approach to forecasting corrosion behavior without the necessity of an extensive use of laboratory testing using chemical-physical technologies, while taking into account the main chemical reactions. Although the model was kept simple, just describing the main reaction and transport processes involved, the mathematical simulations and the related model calibration are in agreement with the laboratory experiments.

\end{document}